\documentclass[pdflatex,sn-mathphys-num]{sn-jnl}

\usepackage{graphicx,color}%
\usepackage{multirow}%
\usepackage{amsmath,amssymb,amsfonts}%
\usepackage{amsthm}%
\usepackage{mathrsfs}%
\usepackage[title]{appendix}%
\usepackage{xcolor}%
\usepackage{textcomp}%
\usepackage{manyfoot}%
\usepackage{booktabs}%
\usepackage{algorithm}%
\usepackage{algorithmicx}%
\usepackage{algpseudocode}%
\usepackage{listings}%
\usepackage{epstopdf}
\usepackage{subfigure}
\usepackage{epsfig}
\usepackage{float}
\usepackage{bm}
\usepackage[T1]{fontenc} 
\usepackage{array}
\usepackage{hyperref}
\usepackage{here}
\graphicspath{{figures/}}
\def\simge{\mathrel{%
       \rlap{\raise 0.511ex \hbox{$>$}}{\lower 0.511ex \hbox{$\sim$}}}}
\def\simle{\mathrel{
       \rlap{\raise 0.511ex \hbox{$<$}}{\lower 0.511ex \hbox{$\sim$}}}}

\newcommand*{\doi}[1]{\href{https://doi.org/#1}{doi:\ #1}}
\newcommand{\figcaption}[1]{\def\@captype{figure}\caption{#1}}
\newcommand{\tblcaption}[1]{\def\@captype{table}\caption{#1}}
\newcommand{\no}{\nonumber}

\newcommand{\vk}{{\bf k}}

\newcommand{\otd}{\tilde{\omega}}
\newcommand{\ktd}{\tilde{k}}

\newcommand{\tautd}{\tilde{\tau}}

\theoremstyle{thmstyleone}%
\theoremstyle{thmstyletwo}%

\theoremstyle{thmstylethree}%

\begin{document}

\title[]{A kernel-derived orthogonal basis for spectral functions from Euclidean correlators}


\author[1,2]{\fnm{Norikazu} \sur{Yamada}}\email{norikazu.yamada@kek.jp}

\affil[1]{\orgdiv{Theory Center, Institute of Particle and Nuclear Studies}, 
\orgname{High Energy Accelerator Research Organization (KEK)}, 
\orgaddress{\street{1-1 Oho}, \city{Tsukuba}, \postcode{305-0801}, \country{Japan}}}

\affil[2]{\orgname{Graduate University for Advanced Studies (SOKENDAI)}, 
\orgaddress{\street{1-1 Oho}, \city{Tsukuba}, \postcode{305-0801}, \country{Japan}}}


\abstract{
Spectral functions play a central role in the characterization of a wide range of physical systems, including strongly interacting quantum field theories and many-body systems. Their non-perturbative determination from Euclidean correlation functions constitutes a well-known ill-posed inverse problem and has motivated the development of numerous reconstruction techniques.
In this work, we propose a systematic, prior-free framework for representing spectral functions using an orthogonal functional basis derived directly from the kernel of Euclidean two-point correlation functions.
We identify a set of lattice-accessible constraints together with the associated basis functions.
These functions can be reorganized into an orthogonal basis within which the spectral function may be approximated in a controlled manner.
Using several model spectral functions, we demonstrate that the proposed expansion captures global spectral features and reproduces low-energy transport coefficients with good accuracy.
While the numerical implementation requires high-precision Euclidean correlator data, the present framework is intended not as a direct reconstruction method, but rather as a tool for extracting robust constraints and overall spectral structures.
The approach may therefore serve as a complementary ingredient or preprocessing step for existing spectral reconstruction techniques.
}


\maketitle

\section{Introduction}\label{sec:intro}
Spectral functions encode essential dynamical information in quantum field theories and many-body systems, including particle excitation spectra, decay rates, and transport coefficients.
In lattice QCD and related numerical approaches, such information is accessed indirectly through Euclidean correlation functions.
The relation between a Euclidean correlator and the corresponding spectral function is typically expressed as an integral transform with a smooth kernel, the inversion of which constitutes a paradigmatic example of an ill-posed problem~\footnote{See, for example, \cite{Epstein:2008} and for review~\cite{Meyer:2011gj}.}.

Numerous techniques have been developed to address this inverse problem, including Bayesian inference and maximum entropy methods~\cite{Gubernatis:1991zz,Jarrell:1996rrw,Sandvik:1998,Nakahara:1999vy,Asakawa:2000tr}, singular value decomposition~\cite{Creffield:1995}, Padé-based approaches~\cite{Vidberg:1977}, the Backus–Gilbert method~\cite{Backus:1968,Press:2007ipz,Hansen:2019idp}, and other regularized reconstruction schemes~\cite{Karsch:1986cq,Jarrell:1989,Cuniberti:2001hm,Umeda:2002vr,Krivenko:2006,Burnier:2011jq}.
While these methods have achieved important progress, they often rely on priors and/or regularization assumptions.
Establishing systematic approaches that make minimal assumptions therefore remains a topic of active interest.
See for recent progress in this and related topics, for example,~\cite{Hansen:2014eka,Hansen:2017mnd,Bulava:2019kbi,Patella:2024cto,Bruno:2024fqc}.

In this work, we explore a complementary direction. Instead of aiming at a direct reconstruction of the spectral function, we focus on extracting a set of constraints and global features that are inherent to the Euclidean correlator itself. From the Euclidean two-point function itself and its derivatives with respect to Euclidean time, we derive a family of relations that constrain weighted averages of the spectral function. 
The weight functions emerging from this construction are fully determined by the kernel of the Euclidean correlator and decay exponentially at large frequencies.

We show that these kernel-derived functions can be reorganized into an orthogonal basis, with respect to which the spectral function divided by frequency can be approximated in a systematic expansion. The expansion coefficients are given directly by the aforementioned Euclidean constraints and do not require the introduction of priors. The resulting approximation is designed to capture the overall structure of spectral functions and, in particular, to provide stable access to low-energy quantities such as transport coefficients.

The framework presented here is primarily methodological in nature.
It is not intended to replace existing spectral reconstruction techniques, but rather to provide an additional analytical tool that can be used to test spectral ans\"atze, generate physically motivated constraints, or supply input to more refined reconstruction methods.
To illustrate its properties, we apply the method to several model spectral functions under idealized conditions and show that the approximation could be even very precise for smooth spectral functions.
A detailed investigation of its numerical performance using realistic lattice data is left for future work.

Related approaches to the present work include both reconstruction-oriented and structure-oriented methods for spectral analysis. Bayesian inference techniques, such as the maximum entropy method and stochastic analytic continuation, aim at reconstructing spectral functions by incorporating prior information and regularization criteria. In contrast, methods based on kernel decomposition, including the Backus–Gilbert approach and Mellin-transform-based expansions, focus on identifying well-conditioned combinations of Euclidean data that can be reliably constrained.

The framework developed here belongs to the latter category. It does not attempt to reconstruct local features of spectral functions but instead provides a kernel-derived orthogonal basis and a set of lattice-accessible constraints. As such, it is naturally suited to serve as a diagnostic tool or as complementary input to more elaborate reconstruction schemes.

The paper is organized as follows.
In sec.~\ref{sec:constraints-basis}, we derive a family of relations emerging from the corresponding Euclidean correlation function and identify the pairs of constraint and basis function.
In sec.~\ref{sec:expression}, a spectral function is expressed as an expansion in the orthogonal basis constructed from the original basis functions.
In sec.~\ref{sec:examples}, our approach is illustrated by introducing a few spectral functions by hand.
Finally, possible applications and potential challenges in the practical implementation are discussed in sec.~\ref{sec:summary}.

\section{Constraints and basis functions}
\label{sec:constraints-basis}

For simplicity, throughout this paper we focus on the Euclidean two-point correlation function of a Hermitian bosonic operator in thermal equilibrium, as typically computed in lattice QCD.
Following the notation of \cite{Laine:2016hma}, the relation between a spectral function $\rho(k^0,\vk)$ and a Euclidean two-point correlation function $\Pi(\tau,\vk)$ in the $\tau$-$\vk$ space is given by
\begin{align}
   \Pi^E(\tau,\vk)
=& {1\over \pi}\int^\infty_{0}\!\!\! dk^0\,\rho(k^0,\vk)\,
   \frac{\cosh\left[\left({\beta\over 2}-\tau\right)k^0\right]}
   {\sinh\left({\beta\over 2}k^0\right)}
\label{eq:baseeq}
\,,
\end{align}
where $(k^0,\vk)$~\footnote{$\omega$ is sometimes used instead of $k^0$.} denotes the momentum vector in Minkowski space and $(k_n,\vk)$ with $k_n=2\pi n T$ is that in Euclidean space, respectively.
$\tau$ varying in $0\le \tau\le\beta$ is the Euclidean time and $\beta=1/T$.

Note that \eqref{eq:baseeq} itself can be regarded as a constraint on weighted averages of the spectral function.
A series of constraints can be derived, for example, by differentiating  \eqref{eq:baseeq} with regard to $\tau$ $n$ times,
\begin{align}
  C_n(\tau,\vk)
=& {1\over Z_n(\tau)}
   \frac{d^n \Pi^E(\tau,\vk)}{d \tau^n}\
= \int_0^\infty\!\!dk^0{\rho(k^0,\vk)\over k^0}S_n(k^0,\tau)
   \qquad (n=0, 1, 2, \cdots)
\label{eq:constraints}
\,.
\end{align}
Hereafter, we focus on the determination of $\rho/k^0$ rather than $\rho$ itself.
$Z_n(\tau)$ is an arbitrary constant and can be freely chosen.
In this work, we choose $Z_n(\tau)$ to normalize the integral of $S_n(k^0,\tau)$ over $k^0$, {\it i.e.}
\begin{align}
& \int_0^\infty\!\!dk^0\,S_n(k^0,\tau)=1
\,.
\end{align}
The functions $S_n(k^0,\tau)$, which emerge from the kernel of the integrand in \eqref{eq:baseeq} through the operation, is given by
\begin{align}
   S_n(k^0,\tau)
=& {s_n(k^0,\tau)\over Z_n(\tau)}
=  {1\over Z_n(\tau)}{k^0\over \pi\sinh\left({\beta\over 2}k^0\right)}
   \frac{d^n}{d\,\tau^n}\cosh\left[\left({\beta\over 2}-\tau\right)k^0\right]
\,.
\label{eq:basis-func}
\end{align}
We call $S_n(k^0,\tau)$ the basis function.
If one chooses $\tau=0$ or $\beta$, the $k^0$ integral in \eqref{eq:constraints} is not guaranteed to be finite.
By eliminating these points, all basis functions decay exponentially in the large $k^0$ region and the integral is made definitely finite.

Note that $C_n(\tau,\vk)$ in \eqref{eq:constraints} can be calculated solely from $\Pi^E(\tau,\vk)$ without any information on $\rho/\omega$.
Therefore, these constraints can be used to test spectral functions obtained using other methods by calculating both sides of \eqref{eq:constraints} if $\Pi^E(\tau,\vk)$ is available.

Before going into details, let us make the above quantities dimensionless for later use.
Supposing that $\rho$ has mass dimension $d_\rho$ and defining the following dimensionless quantities,
\begin{align}
  \ktd^0={\beta k^0\over 2\pi},\ \
  \otd={\beta\omega\over 2\pi},\ \
  \tilde{\vk}={\beta \vk\over 2\pi},\ \
  \tautd={\tau\over\beta}\,,
\end{align}
the relevant quantities in dimensionless unit are written as
\begin{align}
   \tilde{\rho}(\ktd^0,\tilde{\vk})
=& \beta^{d_\rho}\rho(2\pi\ktd^0/\beta,2\pi\tilde{\vk}/\beta)
\,\\
   \tilde{\Pi}^E(\tilde{\tau},\tilde{\vk})
=& \beta^{d_\rho+1}\Pi^E(\beta\tilde{\tau},2\pi\tilde{\vk}/\beta)
\,\\
   \tilde{s}_n(\ktd^0,\tilde{\tau})
=& \beta^{n+1}\,s_n(2\pi k^0/\beta,\tilde{\tau}/\beta)
=  {2\ktd^0\over \sinh\left(\pi\ktd^0\right)} 
   \frac{d^n\,\cosh\left[\left(1-2\tilde{\tau}\right)\pi\ktd^0\right]}{d \tilde{\tau}^n}
\,,\\
   \tilde{Z}_n(\tilde{\tau})
=&  \int_0^\infty\!\!d\ktd^0\tilde{s}_n(\ktd^0,\tilde{\tau})   
\,,\\
   \tilde{S}_n(\ktd^0,\tilde{\tau})
=&  {\tilde{s}_n(\ktd^0,\tilde{\tau})\over \tilde{Z}_n(\tilde{\tau})}
\,,\\
   \tilde{C}_n(\tilde{\tau},\tilde{\vk})
=& \beta^{d_\rho+n+1} C_n(\tilde{\tau}/\beta,2\pi\tilde{\vk}/\beta)
=  {1 \over \tilde{Z}_n(\tilde{\tau})}
   \frac{d^n \tilde{\Pi}^E(\tilde{\tau},\tilde{\vk})}{d \tilde{\tau}^n}
=  \int_0^\infty\!\!d\ktd^0\,
   {\tilde{\rho}(\ktd^0,\tilde{\vk})\over \ktd^0}
   \tilde{S}_n(\ktd^0,\tilde{\tau})
\label{eq:cfunc2}
\,.
\end{align}

As stated above, the values at $\tilde{\tau}=0$ and $1$ are not used to avoid possible ambiguities.
Furthermore, on the lattice it is often difficult to obtain the precise values of $\tilde{\Pi}^E(\tilde{\tau},\tilde{\vk})$ at $\tilde{\tau}=0$ and $1$ because the contact term may contribute to them in nontrivial way.

\begin{figure}[htb]
    \centering
    \includegraphics[width=0.75\linewidth]{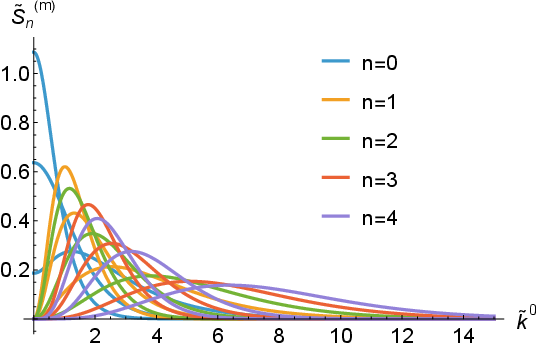}
    \caption{The behaviors of the basis functions, $\tilde{S}_n^{(m)}(\ktd^0)$ for $n=0, 1, 2, 3, 4$ and $\tilde{\tau}=1/8, 2/8, 3/8$.
    Different colors represent different $n$.}
    \label{fig:basisfunc}
\end{figure}

The basis functions $\tilde{S}_n(\ktd^0,\tilde{\tau})$ thus obtained with $n$=0, 1, 2, 3, 4 and $\tilde{\tau}=1/N_T$, $2/N_T$, $3/N_T$, $\cdots$, $(N_T/2-1)/N_T$ are shown in Fig.~\ref{fig:basisfunc} for $N_T=8$.
Although it is not shown in visible way, the peak of each basis function tends to shift to the right and/or to lower as $n$ increases or $\tilde{\tau}$ decreases.

\section{Approximating $\tilde{\rho}/\otd$ in an orthogonal basis}
\label{sec:expression}

We reorganize the original basis $\tilde{S}_i(\ktd^0)$ (defined as $\tilde{S}_n(\ktd^0,\tilde{\tau})$ with $i=(N_T/2-1)n+N_T\tilde{\tau}$) into an orthogonal basis $\tilde{Q}_i(\ktd^0)$,
\begin{align}
&  \tilde{Q}_i(\ktd^0)
=  M_{j,i}\,\tilde{S}_j(\ktd^0)
\label{eq:orth-2}
\,,\\
&  \int_0^\infty\!\!d\ktd^0\,
   \tilde{Q}^*_i(\ktd^0)\,\tilde{Q}_j(\ktd^0)
=  \delta_{i,j}
\label{eq:orth-1}
\,.
\end{align}
Here and thereafter, the sum is taken over repeated indices.
Substituting \eqref{eq:orth-2} into \eqref{eq:orth-1}, we obtain
\begin{align}
&  \int_0^\infty\!\!d\ktd^0\,
   \tilde{Q}_i^*(\ktd^0)\,\tilde{Q}_j(\ktd^0)
=  M_{k,i}^*\,X_{k,l}\,M_{l,j}
=  \delta_{i,j}
\label{eq:orth-3}
\,,
\end{align}
where 
\begin{align}
& X_{k,l}
= \int_0^\infty\!\!d\ktd^0\,\tilde{S}^*_k(\ktd^0)\,\tilde{S}_l(\ktd^0)
\label{eq:matrix-X}
\,.
\end{align}
Since $\tilde{S}_k(\ktd^0)$ is real, $X$ is a real symmetric matrix and can be diagonalized by an orthogonal matrix $O$,
\begin{align}
    O_{j,i}\,X_{j,k}\,O_{k,l}=\lambda_i\delta_{i,l}
\label{eq:diagonalize}
\,,
\end{align}
with real eigenvalues $\lambda_i$.
Then, real matrix $M$ satisfying \eqref{eq:orth-3} is constructed as
\begin{align}
&  M_{i,j}
=  O_{i,k}\,(\lambda_k)^{-1/2}\delta_{k,l}\,O_{j,l}
\,.
\end{align}

We now assume that $\tilde{\rho}/\ktd^0$ can be well approximated by an expansion in the orthogonal basis, namely,
\begin{align}
  \left({\tilde{\rho}(\ktd^0,\tilde{\vk})\over \ktd^0}\right)_{\rm approx}
= \tilde{c}_i(\tilde{\vk})\, \tilde{Q}_i(\ktd^0)
\,,
\end{align}
where the coefficients $\tilde{c}_i(\tilde{\vk})$ are calculated using the orthogonality condition,
\begin{align}
   \tilde{c}_i(\tilde{\vk})
=& \int_0^\infty\!\!d\ktd^0\,
   \left({\tilde{\rho}(\ktd^0,\tilde{\vk})\over \ktd^0}\right)_{\rm approx}\,
   \tilde{Q}_i(\ktd^0)
\no\\
\approx & 
   \int_0^\infty\!\!d\ktd^0\,
   \left({\tilde{\rho}(\ktd^0,\tilde{\vk})\over \ktd^0}\right)\,
   \tilde{Q}_i(\ktd^0)
\no\\
=& M_{j,i}\int_0^\infty\!\!d\ktd^0\,
   \left({\tilde{\rho}(\ktd^0,\tilde{\vk})\over \ktd^0}\right)\,
   \tilde{S}_j(\ktd^0)
\no\\
=& M_{j,i}\,\tilde{C}_j(\tilde{\vk})
\,,
\end{align}
where $\tilde{C}_j(\tilde{\vk})$ corresponds to $\tilde{C}_n(\tilde{\tau},\tilde{\vk})$ defined in \eqref{eq:cfunc2} with $j=(N_T/2-1)n+N_T\tilde{\tau}$.
Combining these results, the approximated $\tilde{\rho}/\ktd^0$ can be expressed as
\begin{align}
  \left({\tilde{\rho}(\ktd^0,\tilde{\vk})\over \ktd^0}\right)_{\rm approx}
=\ & \tilde{C}_j(\tilde{\vk})\,M_{j,i}\, \tilde{Q}_i(\ktd^0)
=  \tilde{C}_j(\tilde{\vk})\,M_{j,i}\, M_{k,i}\,\tilde{S}_k(\ktd^0)
\no\\
=\ & \tilde{C}_j(\tilde{\vk})\,
   O_{j,a}\left(\lambda_a\right)^{-1} O_{k,a}\,
   \tilde{S}_k(\ktd^0)
\label{eq:expansion}
\,,
\end{align}
that is, the spectral function is expressed as an expansion in terms of $\tilde{C}_j(\tilde{\vk})$ and.$\tilde{S}_j(\ktd^0)$.
It can be shown that the approximated spectral function~\eqref{eq:expansion} satisfies the constraint~\eqref{eq:cfunc2}.


As an illustration, the explicit form of the approximation using three basis functions with $n$ = 0 and $N_T = 8$ is given by
\begin{align}
\left({\tilde{\rho}(\ktd^0,\tilde{\vk})\over\ktd^0}\right)_{\rm approx}\
=\ &\
\tilde{C}_0(1/8,\tilde{\vk})\,
\left(  39.461760\,\tilde{S}_0(\ktd^0,1/8)
      - 57.603483\,\tilde{S}_0(\ktd^0,2/8)
      + 29.549685\,\tilde{S}_0(\ktd^0,3/8) \right)
\no\\&
+ \tilde{C}_0(2/8,\tilde{\vk})\,
  \left(- 57.603483\,\tilde{S}_0(\ktd^0,1/8)
        +106.77605 \,\tilde{S}_0(\ktd^0,2/8)
        - 59.773114\,\tilde{S}_0(\ktd^0,3/8) \right)
\no\\&
+ \tilde{C}_0(3/8,\tilde{\vk})\,
  \left(  29.549685\,\tilde{S}_0(\ktd^0,1/8)
        - 59.773114\,\tilde{S}_0(\ktd^0,2/8)
        + 35.741842\,\tilde{S}_0(\ktd^0,3/8) \right)
\,,
\label{eq:master-1st}
\end{align}
where
\begin{align}
     \tilde{S}_0(\ktd^0,1/8)\
=\ & {1\over 2+\sqrt{2}}{2 \ktd^0 \cosh[3 \ktd^0 \pi/4] \over \sinh[\ktd^0 \pi]}
\,,\no\\
     \tilde{S}_0(\ktd^0,2/8)\
=\ & {2 \ktd^0 \cosh[\ktd^0 \pi/2] \over \sinh[\ktd^0 \pi]}
\,,\no\\
      \tilde{S}_0(\ktd^0,3/8)\
=\ & {1\over 2 \sin^2[\pi/8]}{\ktd^0 \cosh[\ktd^0 \pi/4] \over \sinh[\ktd^0 \pi]}
\,.
\end{align}

\begin{figure}
    \centering
    \includegraphics[width=0.75\linewidth]{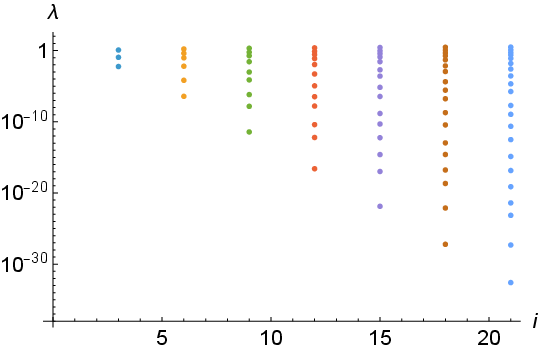}
    \caption{The eigenvalue distribution for the different sizes of matrix $X$ which is $i\times i$.}
    \label{fig:eigenvalue-dist}
\end{figure}

It would be desirable to expand the space spanned by the orthogonal basis.
We attempted to extend the calculation up to $n=6$, but encountered computational difficulties.
As the maximum value of $n$ increases, the size of matrix $X$ defined in \eqref{eq:matrix-X} grows as $3n\times 3n$ in the case of $N_T=8$.
Then, as shown in Figure ~\ref{fig:eigenvalue-dist}, while the largest eigenvalue of $X$ remains in $O(1)$, the smallest eigenvalue decreases exponentially.
This type of behavior commonly occurs in similar situation~\cite{Epstein:2008,Meyer:2011gj} and indicates that including additional terms may not be effective since the presence of the vanishing eigenvalue means that one of the basis is not linearly independent of others and including such basis does not expand the orthogonal basis.
At the same time, since $1/\lambda_a$ appears in \eqref{eq:expansion}, very large intermediate contributions may arise and cancel among themselves, ultimately yielding an $O(1)$ result.
In such cases, both the diagonalization procedure and the constraints must be determined with increasing numerical accuracy in order to control rounding errors.

\section{Examples}
\label{sec:examples}

Three spectral functions are introduced by hand to examine the performance of the approximation.
Since the purpose here is to examine the mathematical aspect of the expansion, an idealized case is assumed, {\it i.e.}, not only the Euclidean correlation function at discrete time points, but also its derivatives, are assumed to be known without uncertainty.

The three spectral function models examined are
\begin{align}
\mbox{model\ 1\ :\ }&
   {\tilde{\rho}_1(\otd,0)\over \otd}
=\ {4\over(2+\otd^2)^2}
   \left[1-{\tanh\left(1-\otd\right)\over 4}{\otd^2\over \ln^2(2+\otd^2)}\right]
\label{eq:modelspect-1}
\,,\\
\mbox{model\ 2\ :\ }&
   {\tilde{\rho_2}(\otd,0)\over\otd}
=\ \left[  {0.3\over 1+(\otd/0.71)^2} 
         + {0.2\over 1+((\otd-7.1)/2.9)^2} + {0.2\over 1+((\otd+7.1)/2.9)^2}
   \right]{2\over 1+(\otd/9.5)^6}
\label{eq:modelspect-2}
\,,\\
\mbox{model\ 3\ :\ }&
   {\tilde{\rho_3}(\otd,0)\over\otd}
=\ {15\left\{1 + 2\,\otd^2 (\otd^2 - {1\over 8}) (\otd^2 - 1) (\otd^2 - {9\over 4}) +
             2\,\otd^2\right\}
   \over 60 + \otd^{12}}
\label{eq:modelspect-3}
\,.
\end{align}
$\tilde{\rho}_1$ is adapted from \cite{Burnier:2011jq}, with a slight modification introduced to ensure that it remains positive, and is inspired by a spectral function associated with an "electric field" correlator yielding the momentum-diffusion coefficient of a heavy quark~\cite{Casalderrey-Solana:2006fio,Caron-Huot:2009ncn,Meyer:2010tt}.
$\tilde{\rho}_2$, inspired by studies of the two-dimensional Hubbard model, is taken from Ref.~\cite{Gunnarsson:2010}, in which the methods based on the maximum entropy, singular value decomposition, sampling, and Pad\'e approximants are compared.
$\tilde{\rho}_3$ is chosen to have a structure qualitatively different from those of $\tilde{\rho}_1$ and $\tilde{\rho}_2$, and is likely to be challenging for many reconstruction methods to reproduce.

Defining the average value of $\ktd^0$ for each basis function as
\begin{align}
   \tilde{\omega}_n(\tilde{\tau})
=& \int_0^\infty\!\!d\ktd^0\, \ktd^0 \tilde{S}_n(\ktd^0,\tilde{\tau})
\end{align}
and calculating the constraints $\tilde{C}_n(\tilde{\tau},0)$ for each of model, we obtain a set of points
$(\otd_n(\tilde{\tau}),\, \tilde{C}_n(\tilde{\tau},0))$.
We emphasize that, under our assumptions, the constraints $\tilde{C}_n$ can be calculated from an accurately determined Euclidean correlation function without knowing $\tilde{\rho}$ in principle.

\begin{figure}[t]
 \centering
 \begin{tabular}{lll}
   \includegraphics[width=0.33\linewidth]{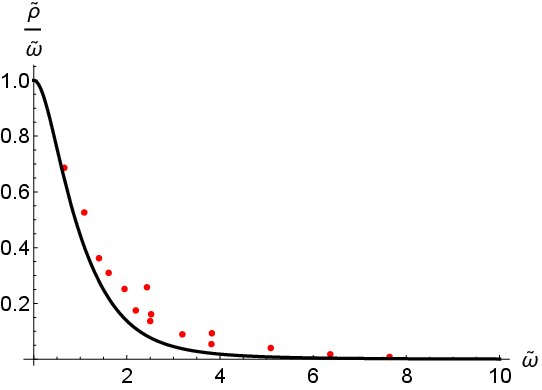}
 & \includegraphics[width=0.33\linewidth]{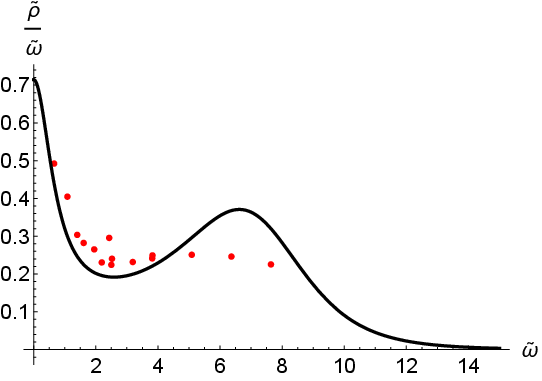}
 & \includegraphics[width=0.33\linewidth]{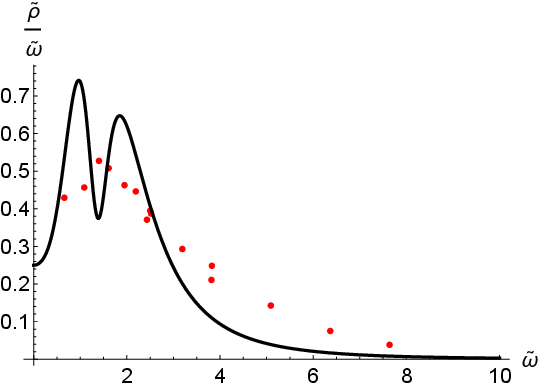}
 \end{tabular}
 \caption{The models for the spectral function (solid curve) and the constraints (red dots) for the models $\tilde{\rho}_1$ (left), $\tilde{\rho}_2$ (middle) and $\tilde{\rho}_3$ (right).
 }
 \label{fig:modelspect}
\end{figure}

In the following, we consider the case with $N_T=8$, then the first three constraints in each model are given by $\tilde{C}_0(\tilde{\tau},0)=\tilde{\Pi}^E(\tilde{\tau},0)/\tilde{Z}_0(\tilde{\tau})$,
\begin{align}
&\mbox{model 1 :}\
\tilde{C}_0(1/8,0)=0.25807411\,,\
\tilde{C}_0(2/8,0)=0.52623779\,,\
\tilde{C}_0(3/8,0)=0.68626502\,,
\label{eq:c-model1}
\\
&\mbox{model 2 :}\
\tilde{C}_0(1/8,0)=0.29559098\,,\
\tilde{C}_0(2,8,0)=0.40467584\,,\
\tilde{C}_0(4/8,0)=0.49234585\,,
\label{eq:c-model2}
\\
&\mbox{model 3 :}\
\tilde{C}_0(1/8,0)=0.37085833\,,\
\tilde{C}_0(2/8,0)=0.45663113\,,\
\tilde{C}_0(3/8,0)=0.42938612\,,
\label{eq:c-model3}
\end{align}
Three sets of data points
$(\otd_n(\tilde{\tau}),\, \tilde{C}_n(\tilde{\tau},0))$ are plotted in Fig.~\ref{fig:modelspect} together with the spectral functions of each.
The data points roughly follow the underlying spectral function in all cases, although the precise functional form cannot be inferred from them alone.

\begin{figure}[t]
 \centering
 \begin{tabular}{cc}
    \includegraphics[width=0.5\linewidth]{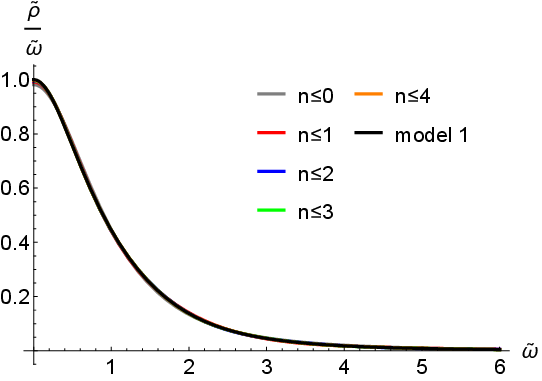}
  & \includegraphics[width=0.5\linewidth]{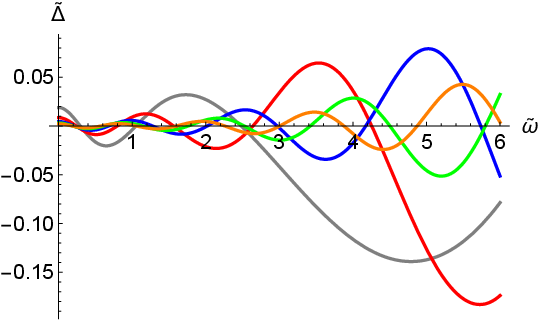}\\
    \includegraphics[width=0.5\linewidth]{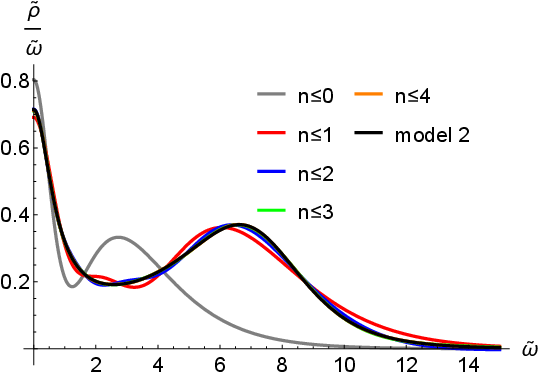}
  & \includegraphics[width=0.5\linewidth]{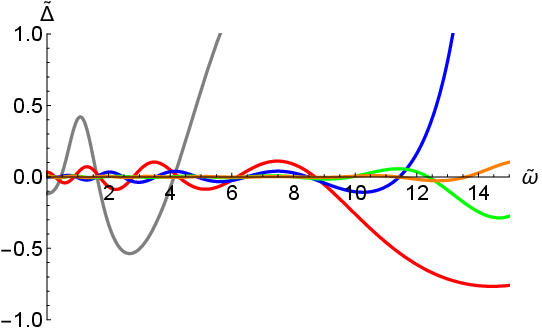}\\
    \includegraphics[width=0.5\linewidth]{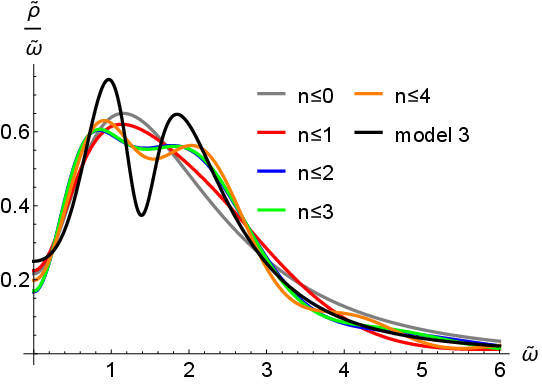}
  & \includegraphics[width=0.5\linewidth]{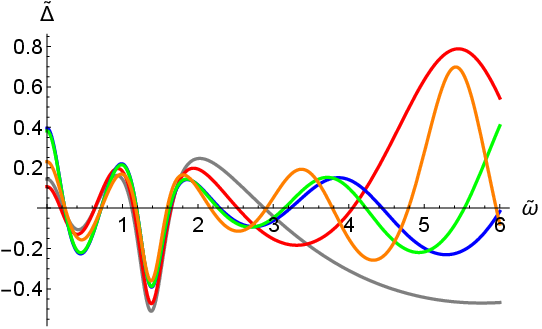}
 \end{tabular}
 \caption{The approximated $\tilde{\rho}/\otd$ and the input (black curve) for the model 1 (top), 2 (middle) and 3 (bottom).}
 \label{fig:modelspect-rslt}
\end{figure}

By substituting the numerical values of the constraints and the basis functions into \eqref{eq:expansion}, we obtain the approximated spectral functions shown in Figs.~\ref{fig:modelspect-rslt},
where we also show the relative difference defined by
\begin{align}
  \tilde{\Delta}
= {{\tilde{\rho}_i\over\otd}-\left({\tilde{\rho}_i\over\otd}\right)_{\rm approx}
  \over
  {{1\over 2}\left[{\tilde{\rho}_i\over\otd}+\left({\tilde{\rho}_i\over\otd}\right)_{\rm approx}\right]}}
\,.
\end{align}

In the case of model 1 (see the top panels of Fig.~\ref{fig:modelspect-rslt}), the approximation successfully reproduces the input spectral function in terms of only three basis functions with $n=0$.
It means that the approximation
\begin{align}
\left({\tilde{\rho}_1(\ktd^0,0)\over\ktd^0}\right)_{\rm approx} 
=0.1498\,\tilde{S}_0(\ktd^0,1/8)
+0.3034\,\tilde{S}_0(\ktd^0,2/8)
+0.6995\,\tilde{S}_0(\ktd^0,3/8)
\label{eq:approx-model1}
\,,
\end{align}
which is obtained by substituting \eqref{eq:c-model1} into \eqref{eq:master-1st}, is already very accurate.
However, it should be noted that each coefficient in \eqref{eq:approx-model1} arises from cancellations among large contributions of $O(10^2)$, and hence the accuracy in the practical calculation will be sensitive to the errors of $\tilde{C}_n(\tilde{\tau},0)$ and the correlation among them.
$\Gamma=\lim_{\otd\to 0}{\tilde{\rho}(\otd)/\otd}$ is often directly related to a transport coefficient and is also well reproduced better than 2 \% in the expansion using only three basis functions.

For model 2 (see the middle panels of Fig.~\ref{fig:modelspect-rslt}), it turns out that including the nine basis functions corresponding to $n\le 2$ yields very successful approximation and $\Gamma$ is reproduced within 12 \%, 4 \%, 0.2 \% for $n=0$, $n\le 1$ and $n\le 2$, respectively.
However, high-precision numerical determinations of $\tilde{C}_n$ are again required to realize these accuracy for $\Gamma$.
It is interesting to compare our result with those based on four different methods adopted in Ref.~\cite{Gunnarsson:2010}.
Although the prerequisites are different and hence the totally fair comparison is not possible, our method using all constraints with $n\le 2$ provides the best approximation.

For the model 3 (see the bottom panels in Fig.~\ref{fig:modelspect-rslt}), even after including 15 basis functions, {\it i.e.} $n\le 4$, the approximation is not as accurate as in the previous two cases, and in particular it fails to reproduce the rapid fluctuation around $\otd=1.5$.
Since the basis functions only contain terms like $(\ktd^0)^n$ or $e^{-\ktd^0\,\pi}$, the approximation will fail to reproduce features with more rapid variations than those encoded in the basis functions.
However, the approximation can be still useful for $\Gamma$.
The constraints with $n=0$, {\it i.e.} $\tilde{C}_0$ , are sensitive to the behavior of the spectral function around $\otd =0$ because the corresponding basis functions $\tilde{S}_0$ has a support around there as seen from Fig.~\ref{fig:basisfunc}.
Thus, $\Gamma$ can be reproduced relatively well in all cases considered.
Indeed, it is reproduced within 13 \% using only $n=0$ constraints for the model 3.

\section{Summary}
\label{sec:summary}

In this work, we have introduced a systematic framework to analyze spectral functions using an orthogonal functional basis derived directly from the kernel of Euclidean two-point correlation functions.
From the Euclidean correlator and its derivatives, we constructed a set of lattice-accessible constraints and associated basis functions that decay exponentially at large frequencies.
These functions can be reorganized into an orthogonal basis within which the spectral function divided by frequency can be approximated in a controlled expansion.

We have illustrated the properties of this expansion using several model spectral functions under idealized conditions.
The results demonstrate that global features of spectral functions, as well as low-energy quantities such as transport coefficients, can be reproduced with good accuracy using only a limited number of basis elements, even when the spectral function exhibits rapid oscillatory behavior.
It is also important to note that the analysis remains unchanged for spectral functions at finite momentum.
At the same time, the analysis reveals inherent numerical challenges arising from the exponentially decreasing eigenvalues, which amplify the sensitivity to statistical and systematic uncertainties in practical applications.

For these reasons, the present framework is not intended to serve as a stand-alone method for the direct reconstruction of detailed spectral functions from lattice data.
Instead, it should be viewed as a diagnostic and preprocessing tool that extracts robust, kernel-determined information from Euclidean correlators. Potential applications include testing spectral ans\"atze, providing physically motivated constraints, or supplying controlled input to more elaborate reconstruction approaches such as Bayesian or regularized inversion methods.

Future work will focus on assessing the numerical performance of this framework using realistic lattice data, including the impact of finite statistics and finite lattice spacing.
Further developments may also explore strategies to improve numerical stability.
We expect that the present approach can provide a useful methodological contribution to the broader toolbox for spectral analyses in lattice QCD and related fields.

\section*{Acknowledgments}
This work is supported in part by JSPS KAKENHI Grant-in-Aid for Scientific Research (No.22K03645).

\end{document}